# A NOT gate in a *cis-trans* photoisomerization model

M. Ndong[1,2], L. Bomble[1,2], D. Sugny[3], Y. Justum[1,2] and M. Desouter-Lecomte[1,2,4] *

[1] Univ Paris-Sud, Laboratoire de Chimie Physique, UMR8000, Orsay, F-91405.

[2] CNRS, Laboratoire de Chimie Physique, UMR8000, Orsay, F-91405.

[3] Institut Carnot de Bourgogne, Unité Mixte de Recherches 5027 CNRS et Université de Bourgogne, BP 47870, 21078 Dijon, France.

[4] Département de Chimie, Université de Liège, Institut de Chimie B6, Sart-Tilman, B-4000, Liège 1, Belgium.

* corresponding author: mdesoute@lcp.u-psud.fr

**Abstract**

We numerically study the implementation of a NOT gate by laser pulses in a model molecular system presenting two electronic surfaces coupled by non adiabatic interactions. The two states of the bit are the fundamental states of the *cis-trans* isomers of the molecule. The gate is classical in the sense that it involves a one-qubit flip so that the encoding of the outputs is based on population analysis which does not take the phases into account. This gate can also be viewed as a double photo-switch process with the property that the same electric field controls the two isomerizations. As an example, we consider one-dimensional cuts in a model of the retinal in rhodopsin already proposed in the literature. The laser pulses are computed by the Multi Target Optimal Control Theory with chirped pulses as trial fields. Very high fidelities are obtained. We also examine the stability of the control when the system is coupled to a bath of oscillators modelled by an Ohmic spectral density. The bath correlation time scale being smaller than the pulse duration the dynamics is carried out in the Markovian approximation

# 1. Introduction

Manipulating quantum systems by using time-dependent electric field remains a goal of primary interest in different molecular processes extending from the control of chemical reactions [1,2] to quantum computing [3]. According to the degrees of freedom involved in the control i.e. rotational, vibrational or electronic, the processes considered are different. Among these, we can cite molecular alignment and orientation [4,5,6], isomerization by vibrational excitations [7,8,9,10] and isomerization by non-adiabatic electronic transitions [11,12,13] which have been the subject of a large amount of theoretical works. The control fields have been determined by different control schemes such as the coherent control [14,15], the local control approach [16,17,18,19] and the Optimal Control Theory (OCT) [20,21,22] or by adiabatic processes [23,24,25] when the system is sufficiently simple or possesses particular symmetries. This paper focuses on non-adiabatic electronic transitions. The possibility to control the photoisomerization process has been recently shown experimentally for the 3,3'-diethyl-2,2'-thiacyanine iodide (cyanine dye NK88) [26] and to some extend for the chromophore of the rhodopsin [27] illustrating the fact that wave properties can be observed and manipulated even in very complex systems. The mechanism of control of NK88 has been studied theoretically in a simplified model consisting of a one degree of freedom system coupled to a bath [28]. A quantitative agreement with the experimental results has been obtained. Even if the role and the influence of the other molecular degrees of freedom are still discussed in these systems [29], this latter work shows that simple models are not unrealistic and can help understanding the structure of the control.

In view of these studies, a question which naturally arises is the control of more complex reactions in these systems. We investigate here the control by a single laser pulse of the *double photoisomerization* process or, in other words, of the *double photo-switch*. The goal of the control is to steer the system from the fundamental vibrational state of the isomer *cis* to the fundamental vibrational state of the isomer *trans* and *vice versa* with the same electric field.

This precisely corresponds to a NOT logical gate in a two-state system. Note that our objective is more challenging than just a double photoizomerization defined from the population of the electronic states. Implementing logical gates on molecular systems is based on a classical logical approach [30,31,32,33] or on quantum computing. In the latter case, the qubits have been encoded in rotational levels [34], vibrational normal modes [35,36,37,38,39,40,41,42], ro-vibrational states [43] and ro-vibrational states belonging to different electronic surfaces [44,45]. The gate operations are realized by laser pulses. A possible choice for a molecule with two isomers *cis-trans* is to define a bit or a qubit from the vibrational ground states of the two minima of the diabatic potential energy surfaces. However, up to date, little has been done for implementing gates defined from isomers involving nonadiabatic interactions. This is basically due to the difficulty of the control which involves a large number of quantum levels and potential energy crossings [11,12,46,47,48,49, 50].

This double photoisomerization control is particularly challenging when the two isomers do not play a symmetrical role. The laser pulse realizing the gate is then expected to be slightly different from the laser field controlling the photoisomerization. Due to the complexity of the control, we consider only a *classical not* gate i.e. the encoding of the outputs is based on population analysis which does not take the phases into account [41,51,52]. As a first test of feasibility, we consider a model of the retinal in rhodopsin already proposed in the literature [53,54] and used in different works [11,12,50]. This is a very simplified model even if recent theoretical investigations have emphasized the importance of the multidimensionality for photo-physics with conical intersections [55,56]. In the spirit of the simulation on the cyanine dye control [28], we first consider the dominant isomerization coordinate which is a torsion angle denoted $\phi$. For more realistic applications including more active degrees of freedom, it will be possible to use the promising OCT-MCTDH (Multiconfiguration Time-Dependent Hartree) method [57]. Then we couple this active coordinate $\phi$ with a bath of oscillators described by an Ohmic spectral density as it has been frequently used in OCT simulations [28, 58,59,60,61] and carry out dissipative Markovian dynamics because we choose a bath with a small correlation time compared to the pulse duration. Non Markovian dynamics could be considered [62,63,64,65,61,66] but at the price of a very long computation time in this example. For short pulses and complex systems in which the timescales cannot be separated the Surrogate Hamiltonian method represents an interesting alternative to address quantum dissipative dynamics [67,68,69,70].

We determine the control fields by the Multi target Optimal Control Theory [35] which provides an optimal universal field able to steer the system from a set of initial states to a set of target states. We observe the crucial role of the trial field in the successful application of this control strategy. We use here chirped laser pulses as trial fields. Several works have already pointed out the efficiency of such electric fields in the control of non-adiabatic dynamics [71,72,73,74].

This paper is organized as follows. In Sec. 2, we introduce the model Hamiltonian and we recall the different steps of multi-target OCT. The control scheme is then applied to the retinal in Sec. 3. We discuss the qualitative characteristics of the optimal pulse in each case and its robustness with respect to the dissipation. Conclusions and prospective views are given in Sec. 4.

## 2. Model and methodology

*2.1. Model Hamiltonian*

We consider cuts in a two-dimensional model of the retinal built to reproduce efficiently the time resolved emission [53,54]. The model includes two electronic surfaces with a conical intersection. The active degree of freedom is here the large amplitude torsional mode $\phi$ which is by definition periodic. The second coordinate of the initial model is an effective coupling mode *x* which roughly corresponds to a stretching mode of the polyene chain. The reduced one-dimensional Hamiltonian matrix **H** of the system can be written in the diabatic electronic basis set as

$$\mathbf{H} = \mathbf{H}_0 - \vec{\mathbf{\mu}}.\vec{E}(t) \tag{1}$$

where

$$\mathbf{H}_0 = \mathbf{T} + \mathbf{V} = \left[-\frac{\hbar^2}{2I}\frac{\partial^2}{\partial \phi^2}\right]\mathbf{1} + \begin{bmatrix} V_{11} & V_{12} \\ V_{21} & V_{22} \end{bmatrix} \tag{2}$$

is the field-free Hamiltonian, $\vec{\mathbf{\mu}}$ the dipole operator and $\vec{E}(t)$ the electric field which is linearly polarized. We assume that the dipole operator has non-zero matrix elements only between states belonging to two different diabatic electronic surfaces ($\mu_{12} = \mu_{21} = 1D$). The parameters of the diabatic electronic basis set $V_{jk}$, the inertia momentum *I* and the mass *m* are given in Ref. [54]. The 1D periodic model corresponds to two different cuts at $x_{(1)}$ = 0.715 bohr and $x_{(2)}$ = 1.43 bohr for which the electronic couplings are respectively $V_{12}^{(1)}$ = 0.005 hartree and $V_{12}^{(2)}$ = 0.01 hartree. The diabatic curves of the model are given in Fig.1.

*2.2. Optimal Control Theory*

The universal field of the gate is computed by the multi-target extension of the optimal control theory [35,36,37,40,41]. The objective is to find the field able to drive each of the $2^N$ initial states of a *N*-qubit system towards the corresponding final states given by the gate unitary transformation

$$\chi_n^{output} = \mathbf{U}_{gate}\chi_n^{input} \tag{3}$$

The functional can be defined in different manners [20,21] which are strongly related [22]. We choose the functional which decouples the boundary conditions [20] for the initial wave packet and the Lagrange multiplier. This functional reads [35,40,41]

$$J = \sum_{n=1}^{2^N} \left\{ \left| \langle \psi_i^n(t_f) | \chi_n^{output} \rangle \right|^2 - 2\Re\left[ \int_0^{t_f} \langle \psi_i^n(t) | \psi_f^n(t) \rangle \langle \psi_f^n(t) | \partial_t + \frac{i}{\hbar} \hat{H} | \psi_i^n(t) \rangle dt \right] \right\} - \alpha \int_0^{t_f} E^2(t) dt$$

(4)

where $N$ is the number of qubits (here $N = 1$), $t_f$ is the duration of the pulse and $\alpha$ is a positive penalty factor which limits the laser fluence. $\psi_i^n(t)$ is the $n^{\text{th}}$ wave packet propagated forwards with the optimal field $E(t)$ with initial value $\psi_i^n(t=0) = \chi_n^{input}$. $\psi_f^n(t)$ is the Lagrange multiplier ensuring that the Schödinger equation is satisfied at any time. $\psi_f^n(t)$ is propagated backwards with the final condition $\psi_f^n(t_f) = \chi_n^{output}$. $\left| \langle \psi_i^n(t_f) | \chi_n^{output} \rangle \right|^2$ is the performance index of the $n^{\text{th}}$ transformation and the fidelity of the gate is given by

$$F = \frac{1}{2^N} \sum_{n=1}^{2^N} \left| \langle \psi_i^n(t_f) | \chi_n^{output} \rangle \right|^2 \tag{5}$$

The optimal field is finally expressed as a sum over all the transformations of the gate

$$E(t) = -\frac{s(t)}{\hbar \alpha} \Im m \sum_{n=1}^{2^N} \left[ \langle \psi_i^n(t) | \psi_f^n(t) \rangle \langle \psi_f^n(t) | \mu | \psi_i^n(t) \rangle \right] \tag{6}$$

where the envelope $s(t) = \sin^2(\pi t / t_f)$ has been introduced to induce a smooth in and off [75]. The time evolution is carried out by the split operator method [76] extended to non adiabatic processes [77]. The elementary evolution operator for a time step is given by

$$U(\delta t)\psi(t_k) = e^{-i\frac{\delta t}{4\hbar}\mathbf{V}} e^{i\frac{\delta t}{2\hbar}\mu E(t_k)} e^{-i\frac{\delta t}{4\hbar}\mathbf{V}} e^{-i\frac{\delta t}{\hbar}\mathbf{T}} e^{-i\frac{\delta t}{4\hbar}\mathbf{V}} e^{i\frac{\delta t}{2\hbar}\mu E(t_k)} e^{-i\frac{\delta t}{4\hbar}\mathbf{V}} \psi(t_k) \tag{7}$$

We adopt the iteration scheme of Ref. [20] and we use the improvement proposed in Ref. [44] in order to speed up the convergence of the algorithm. At each iteration, the field is given by $E^{(k)} = E^{(k-1)} + \Delta E^{(k)}$ where $\Delta E^{(k)}$ is calculated by Eq. (6). The spatial grid contains $2^{10}$ points and the time step is 0.024 fs.

The environment is introduced by coupling the system to a dissipative bath which is composed of a set of $N_b$ harmonic oscillators $Q_j$. The system-bath coupling is given by $\hat{H}_{SB} = -f(\phi) \sum_{j}^{N_b} c_j Q_j$ where the operator $f(\phi)$ is a diagonal matrix in the diabatic basis with $f(\phi) = \cos(\phi) + \sin(\phi)$ on the diagonal. Note that this latter choice does not imply particular symmetry in the coupling. The spectral density of the bath $J(\omega) = (\pi/2) \sum_{j}^{N_b} (c_j^2 / \omega_j) \delta(\omega - \omega_j)$ with $J(-\omega) = -J(\omega)$ is approximated by an Ohmic function [78]

$$J(\omega) = \lambda^2 (\omega / \omega_c) \exp-(|\omega|/\omega_c). \tag{8}$$

We choose $\omega_c = 400$ cm$^{-1}$ (a similar value of 450 cm$^{-1}$ is taken in ref. [28]) and $T = 300$ K. The relaxation time $\tau_R$ is of the order of $1/\lambda^2$. When $\lambda$ varies from $\lambda = 10^{-3}$ to $5 \cdot 10^{-3}$, $\tau_R$ varies from about 25 ps to 1 ps. The time scale $\tau_B$ of the bath dynamics is fixed by $\omega_c$ and the temperature $T$. $\tau_B$ is here of the order of 10 fs for $T = 300$K and is thus smaller than both the pulse duration ($t_f = 500$ fs) and the relaxation time. The Markovian approximation is therefore justified [79]. The density matrix $\boldsymbol{\rho}$ expressed in the electronic diabatic representation can be written as follows

$$\boldsymbol{\rho} = \begin{bmatrix} \rho^{11} & \rho^{12} \\ \rho^{21} & \rho^{22} \end{bmatrix}.$$

$\boldsymbol{\rho}$ is first expressed in the basis set of $N_1$ and $N_2$ vibrational eigenstates of the two diabatic wells, with $N_1 = N_2 = 250$. The $\mathbf{H}$ matrix is then diagonalized in order to use the Lindblad equation [80,81] which is given in the eigenbasis set of the Hamiltonian [Eq. (1)]. Without dissipation, the density matrix evolves according to the Liouville equation $\dot{\boldsymbol{\rho}} = -\frac{i}{\hbar}[\mathbf{H}, \boldsymbol{\rho}]$. The dissipative part takes the form

$$\begin{aligned}\dot{\rho}_{kl} &= -(1/2) \sum_{m=1}^{N_1+N_2} [\gamma(\omega_{mk})|A_{mk}|^2 + \gamma(\omega_{ml})|A_{ml}|^2] \rho_{kl} \\ \dot{\rho}_{kk} &= \sum_{m=1}^{N_1+N_2} \left[\gamma(\omega_{km})|A_{km}|^2 \rho_{mm} - \gamma(\omega_{mk})|A_{mk}|^2 \rho_{kk}\right]\end{aligned} \tag{9}$$

where $\omega_{mk} = (\varepsilon_k - \varepsilon_m)/\hbar$, $\gamma(\omega) = J(\omega)/(1-e^{-\beta\omega})$, $\beta = 1/kT$ and $\mathbf{A}$ is a two by two matrix containing on the diagonal the matrices $A_{mk}$ of the coupling function $f(\phi)$.

## 3. Results

The two states of the bit are the two vibrational ground states of the diabatic electronic states corresponding to the two isomers *cis* and *trans*. These states are denoted by $|0\rangle = \chi_0^{cis}$ and $|1\rangle = \chi_0^{trans}$. The optimal laser field drives the system from the ground vibrational state of the *cis* potential to the ground vibrational state of the *trans* potential and *vice versa*. This can be summarized by the following diagram

$$\text{NOT}|0\rangle = |1\rangle \qquad (10)$$

$$\text{NOT}|1\rangle = |0\rangle. \qquad (11)$$

We first detail the strategies used to obtain optimal fields. We have began by optimizing a single transformation $\chi_0^{cis} \to \chi_0^{trans}$ and we have chosen the corresponding optimal field as a trial field to optimize the NOT gate. The trial fields for the first optimization are chirped pulses of the form [71,72]

$$E^{(0)}(t) = E_{max} e^{-\frac{(t-t_m)^2}{2\sigma^2}} \cos[\omega(t)(t-t_m) + \varphi] \qquad (12)$$

with $\omega(t) = \omega_0 + c(t-t_m)$. We have used a short chirp $E_1^{(0)}(t)$ leading to a Franck Condon transition followed by a longer second chirp $E_2^{(0)}(t)$ for the rest of the control. The parameters are gathered in Table I. They are selected because they give the best performance index at the first iteration (at least of the order of $10^{-3}$).

Table I. Parameters of the chirped pulses [Eq. (12)] used as trial fields

| Chirp | $E_{max}$ (Vm$^{-1}$) | $t_m$ (fs) | $\sigma$ (fs) | $\hbar\omega_0$ (cm$^{-1}$) | $\hbar c$ (cm$^{-1}$/ps) | $\varphi$ |
|---|---|---|---|---|---|---|
| $E_1^{(0)}(t)$ | 5.91 10$^9$ | 12 | 3.4 | 21 945 | 548.6 | 0 |
| $E_2^{(0)}(t)$ | 8.06 10$^8$ | 230 | 65 | 13 123 | 17.1 | 0 |

*3. 1. Control without dissipation*

The results are illustrated for the case $V_{12} = 0.01$ hartree. Figure 2 shows the evolution of the population of the two electronic diabatic states for the two transformations with the field that optimizes only the $\chi_0^{cis} \to \chi_0^{trans}$ isomerization. This illustrates the fact that the optimal field for the *cis-trans* transformation is not directly able to perform the NOT gate. The first performance index of the reverse *trans-cis* process is of the order of 0.1 %. Figure 3 gives the population evolution for the gate field. One observes the expected population inversion. However, this global information must be completed by the value of the performance index to assess that the final wave packet is effectively cooled towards the ground vibrational state. One obtains a performance index of 96.9% for the transformation $\text{NOT}|0\rangle = |1\rangle$ and 96.1% for $\text{NOT}|1\rangle = |0\rangle$. The mechanism is slightly different for the two transformations. For example, one observes the sharp Franck-Condon jump induced by the first chirp $E_1^{(0)}(t)$ for t < 0.05 ps in the $\text{NOT}|0\rangle = |1\rangle$ (*cis-trans*) case while the final jump is not so sharp at the end of the reverse transformation (*trans-cis*) for t > 0.45 ps.

Fig. 4 gives the optimal field of the single *cis-trans* izomerisation (upper part) and of the NOT gate (lower part). The second field is more complex. Fig. 5 displays the Gabor transforms of these two fields, the upper panel corresponds to the simple *cis-trans* isomerization and the lower panel to the NOT gate. The Gabor transform is defined by

$$F(\omega,t) = \left| \int_{-\infty}^{+\infty} H(s-t,\tau)E(s)e^{i\omega s}ds \right|^2 \tag{13}$$

where $H(s,\tau)$ is the Blackman window [82] and

$$H(s,\tau) = 0.08\cos(\frac{4\pi}{\tau}s) + 0.5\cos(\frac{2\pi}{\tau}s) + 0.42 \text{ if } |s| \leq \frac{\tau}{2}$$
$$H(s,\tau) = 0 \text{ elsewhere,}$$

$\tau$ is the time-resolution fixed here at $\tau$ = 12 fs. The trial field ($E_1^{(0)}(t) + E_2^{(0)}(t)$) is superimposed in dotted lines in the upper part of Fig. 5. The main frequencies used for the control after the Franck Condon jump are those offered by $E_2^{(0)}(t)$ (frequencies of the order of 13 200 cm$^{-1}$ which corresponds to the difference between the diabatic minima). The optimization lets appear new low frequencies (around 8 000 cm$^{-1}$) at early times. They can be related to transitions after the Franck Condon jump leading to nearly equally populated states. Small population exchanges occur up to the cooling when the wave packet is finally localized in the bottom of the *trans* well. The Gabor transform of the NOT field (lower part of Fig. 5) shows that this field has more low frequencies (around 13 000 cm$^{-1}$). These frequencies give at early times the same populations for the two electronic states which is characteristic of the *trans-cis* pathway (see Fig. 3). The behaviour is confirmed by the evolution of the mean energy $\langle \psi_i^n(t) | H_0 | \psi_i^n(t) \rangle$, for $n$ = 1 (*cis-trans*) drawn in Fig. 6. The mean energy is of the order of 0.1 hartree after 0.02 ps. Some exchanges of population are observed during the intermediary time and lead to a very small variation of the average energy up to the final cooling.

Table II gathers the performance indexes for two examples with diabatic couplings $V_{12}$ = 0.01 hartree and $V_{12}$ = 0.005 hartree. We keep the same zero-order trial field ($E_1^{(0)}(t)$ $+ E_2^{(0)}(t)$). The behavior of the electronic population remains roughly the same. No special feature appears due to the different value of the coupling.

*3. 2. Control with dissipation*

We have carried out a controlled dynamics with dissipation [Eq. (9)] for two coupling strengths $\lambda = 10^{-3}$ and $\lambda = 5 \ 10^{-3}$ [Eq. (8)] with a reference frequency $\omega_c = 400$ cm$^{-1}$ and a bath temperature $T = 300$ K. The performance index of a transformation is given by

$$F_{dis} = \frac{1}{2^N} \sum_{n=1}^{2^N} Tr\left[\mathbf{W}_n \boldsymbol{\rho}_n(t_f)\right] \tag{14}$$

where $\mathbf{W}_n$ is the target density matrix for the $n^{th}$ transformation of the gate and $\boldsymbol{\rho}_n(t_f)$ the final density matrix propagated with the optimal field. The initial matrices are those of pure states corresponding to the $|0\rangle$ and $|1\rangle$ states. The trial field is the field optimized without dissipation. We have observed that an optimization with Markovian dynamics does not modify significantly the optimal field. In other words, no new pathway is found by the algorithm in presence of dissipation. The performance index decreases smoothly as the coupling increases but the general behaviour remains the same. This is probably related to the short duration of the pulse compared to the relaxation time ($\tau_R \approx 25$ ps for $\lambda = 10^{-3}$ and 1 ps for 5 $10^{-3}$). Similar results have already been obtained in different adiabatic cases [83,84]. This is in agreement with recent systematic analysis showing that the control cannot completely cancel the effect of dissipation for a dynamics governed by the Lindblad equation [85,86]. However, we observe that laser driven dynamics fights against the effect of dissipation in the sense that the optimal field limits the decoherence due to field-free dissipation. This is illustrated in Fig. 7 where we compare $Tr[\boldsymbol{\rho}^2]$ for a field-free evolution of a Franck Condon wave packet prepared in the excited state for the case $V_{12} = 0.01$ hartree and $Tr[\boldsymbol{\rho}^2]$ of the laser driven process for the *cis-* *trans* transformation. We choose a Franck-Condon wave packet because the initial ground state of the *cis*-well state is quasi stationary and does not lead to non adiabatic dynamics. It is seen that the decrease of $Tr[\boldsymbol{\rho}^2]$ is larger in the field-free case. A similar improvement of the coherence with control in comparison with field-free evolution has been shown in ref. [87] for a completely different model. We can conclude that the control scenarios are quite robust against a limited dissipation. This also means that although laser control cannot completely cancel dissipative effects, high fidelities can still be obtained. This result is finally encouraging for future works taking into account more degrees of freedom of the system. The dissipation plays here the role of a very large number of these degrees of freedom and is the most unfavourable situation. Fields optimized by coupling the system with few modes using the coupled channels [66,84] or the surrogate Hamiltonian [67,68,69,70] could probably give higher fidelities.

Table II. Fidelity of the NOT gate [Eqs. (5) and (14)] without ($\lambda = 0$) and with Markovian dissipation. $\lambda$ fixes the strength of the coupling to the surrounding [Eq. (8)], $\omega_c = 400$ cm$^{-1}$.

| $V_{12}$ (hartree) | Performance index [Eq.(5)] | | |
|---|---|---|---|
| | $\lambda = 0$ | $\lambda = 10^{-3}$ | $\lambda = 5\ 10^{-3}$ |
| 0.01 | 0.965 | 0.941 | 0.806 |
| 0.005 | 0.961 | 0.938 | 0.805 |

## 4. Concluding remarks

The implementation of a NOT gate or double photo-switch in a sub-pico time scale is very appealing. We simulate here a one-dimensional model which may seem rather unrealistic.

However, such a model is already very demanding to achieve a solution to this control problem. The feasibility of such a control must be taken as a first encouraging step before undertaking more complex simulations. The logical gate has been realized by laser pulses determined by OCT. Good results have been obtained since in each example the fidelity is larger than 95%.

Due to the difficulty of the control, the choice of the trial field is particularly crucial. From a numerical point of view, we also point out that the choice of chirp pulses as trial fields has been the only way to reach the convergence of the algorithm. As could be expected, the effect of the coupling to an environment does not drastically modify the result of the control. We have observed a smooth decrease of the efficiency of the control as the effect of dissipation increases but no new pathway is created by the algorithm.

In the scheme we have proposed, only the population has been used to define the target of the control which renders the corresponding gate classical in nature. A first question is the realization of other gates which also involve population flip. An example is ~~given by~~ the basic CNOT (controlled-NOT) gate. The CNOT gate requires however the definition of the second bit. A solution could be to take into account other electronic surfaces in the same molecule or other degrees of freedom (vibrational or rotational). As the phase is also at our disposal, another open question is the generalization of the present study to quantum logical operations involving superposed states such as the Hadamard gate. This seems a difficult task due to the complexity of the system.

We have considered in this paper a model of the retinal but the results obtained are expected to be transposable to other molecules which are characterized by qualitatively similar potential energy curves along the isomerization path. An example of this class of molecules is given by photo-switching molecules such as the spiropyran [88]. Finally, we notice that the experimental realization of such processes seems possible and could be made in a near future since the control of photo-isomerization has already been achieved by adaptative femtosecond pulse shaping [26].

**Acknowledgments**

The computing facilities of IDRIS (Project numbers 061247 and 2006 0811429) as well the financial support of the FNRS in the University of Liège SGI Nic and Nic2 projects are gratefully acknowledged.

**Figure captions**

FIG. 1. Diabatic potential energy curves of the 1D retinal model.

FIG. 2. Upper part: population of the two diabatic electronic states during the evolution with the field optimized only for the *cis-trans* isomerization. $V_{12}$ = 0.01 hartree and the trial field $E_1^{(0)} + E_2^{(0)}$ (see Table I). Lower part: population of the two diabatic electronic states starting from the *trans* isomer with the same optimized field.

FIG. 3. Population of the two diabatic electronic states for the two transformations of the NOT gate for $V_{12}$ = 0.01 hartree. Upper part $\text{NOT}|0\rangle = |1\rangle$ (*cis-trans*), lower part $\text{NOT}|1\rangle = |0\rangle$ (*trans-cis*). The trial field is the field optimized for the single *cis-trans* isomerization used in Fig. 2.

FIG. 4. Optimal field for $V_{12}$ = 0.01 hartree. Upper panel: transformation *cis-trans* with the trial field $E_1^{(0)} + E_2^{(0)}$; lower panel: NOT gate with the upper field as trial field.

FIG. 5. Gabor transform of the optimal fields of Fig.4. Upper panel: transformation *cis-trans*, lower panel: NOT gate.

FIG. 6. Average energy $\langle \psi_i^n(t)|H_0|\psi_i^n(t)\rangle$ during the two transformations of the NOT gate with $V_{12}$ = 0.01 hartree. The full and dashed lines correspond respectively to *n* = 1 and the *cis-trans* transformation and to *n* = 2 and the *trans-cis* transformation.

FIG 7. $Tr[\rho^2(t)]$ for the field-free evolution of a Franck Condon wave packet (gray line) and for the laser driven *cis-trans* isomerization in the case $V_{12} = 0.01$ hartree (black line).

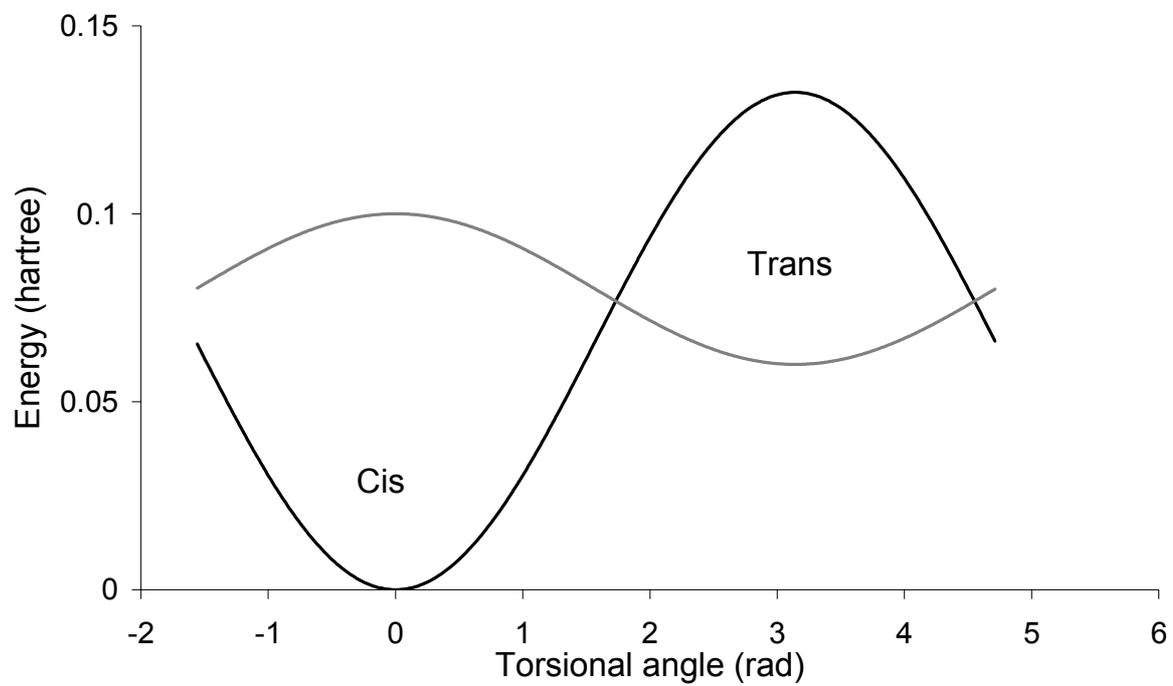

FIG. 1

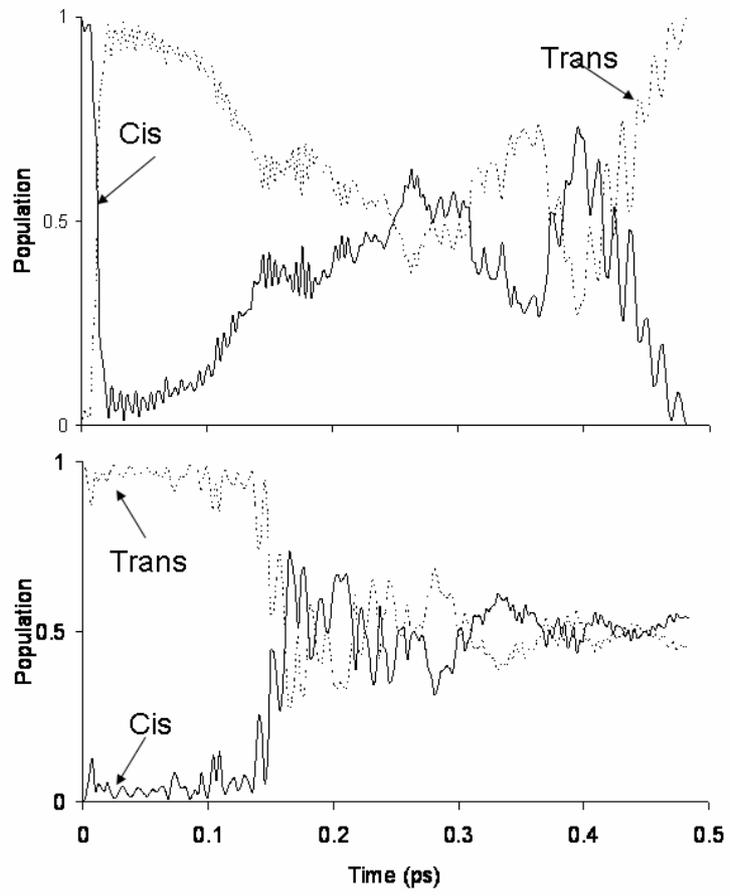

FIG. 2

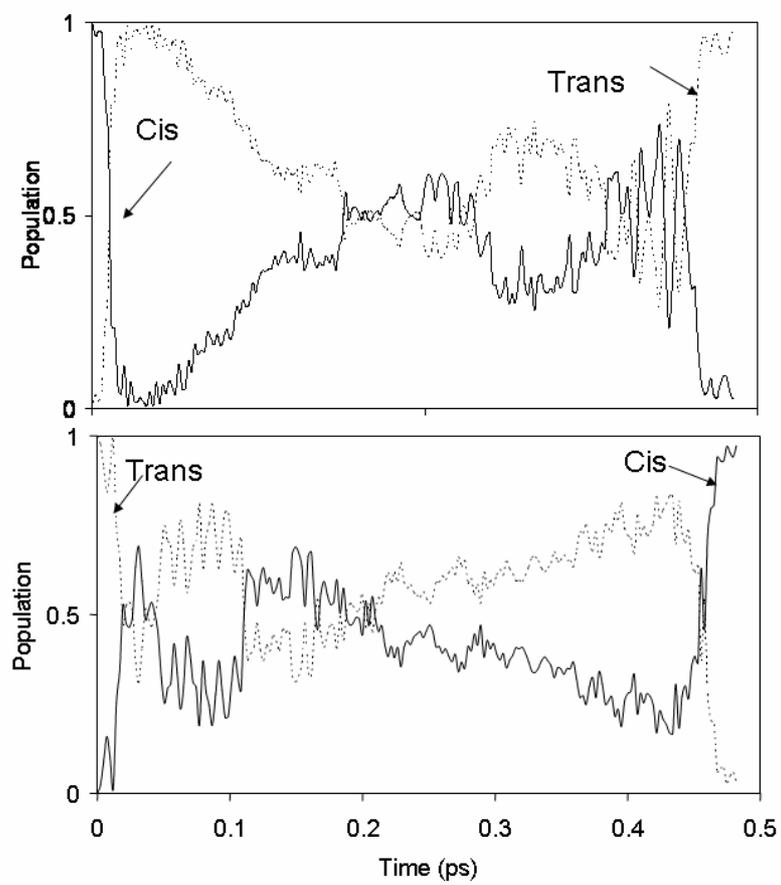

FIG. 3

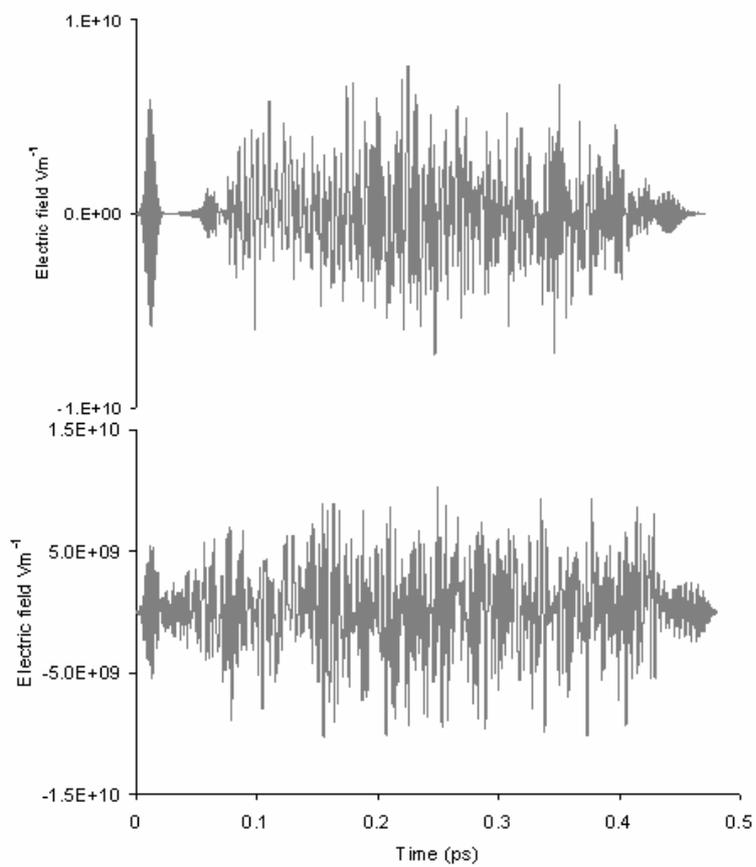

FIG. 4

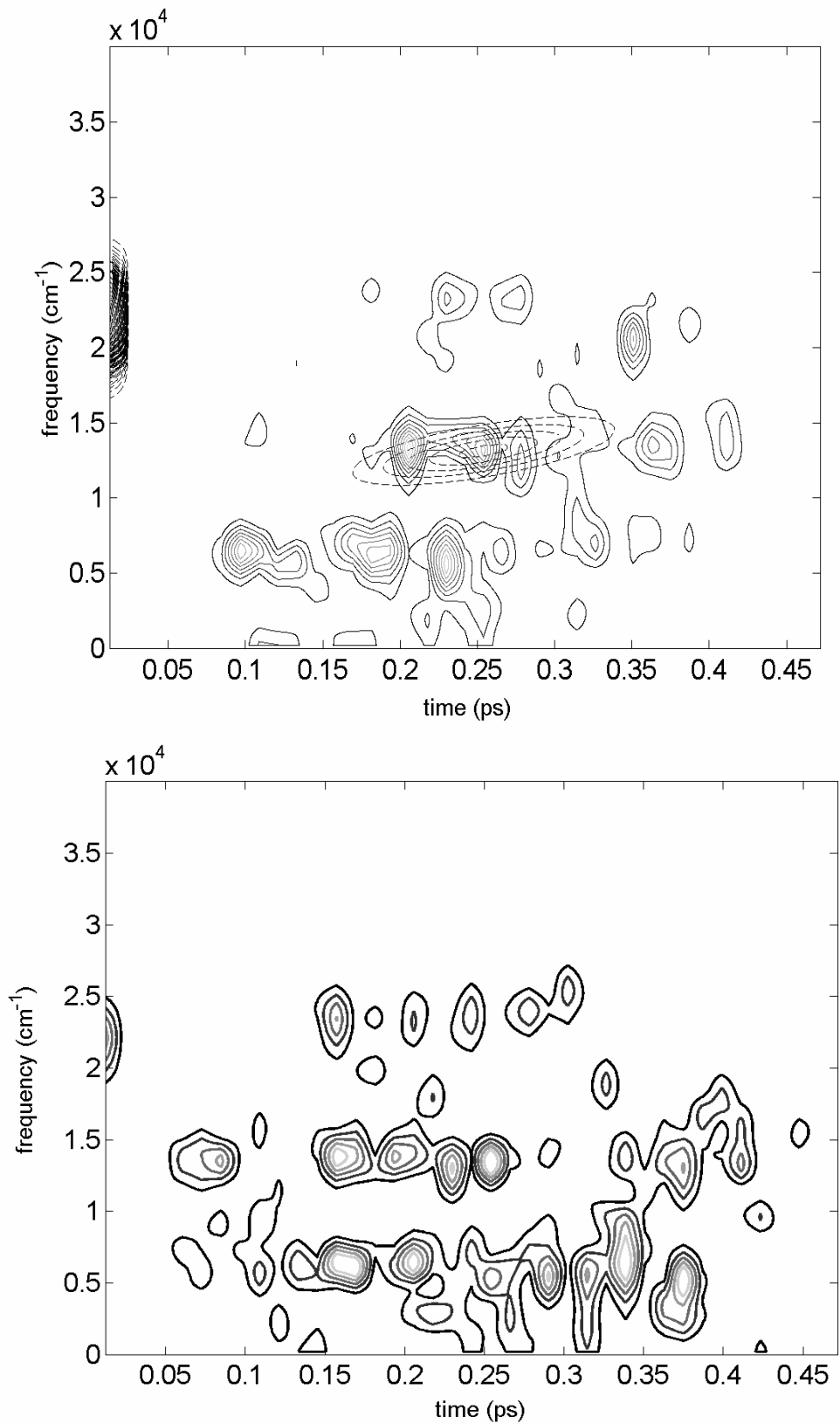

FIG. 5

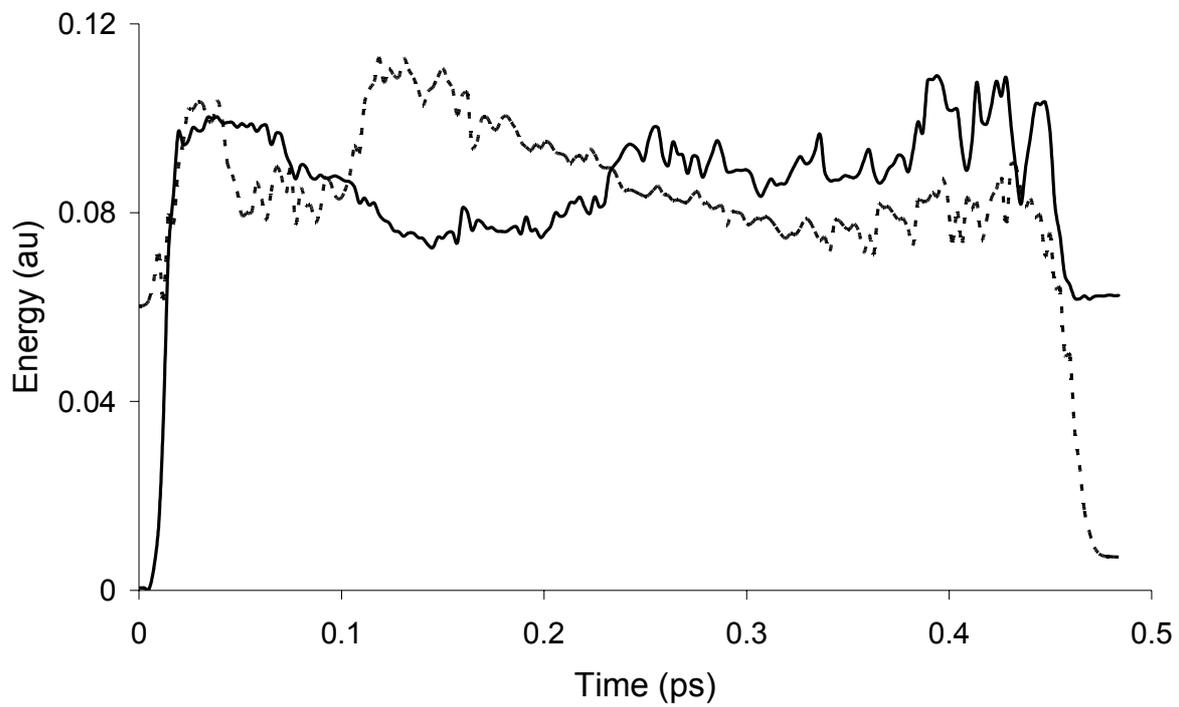

FIG. 6

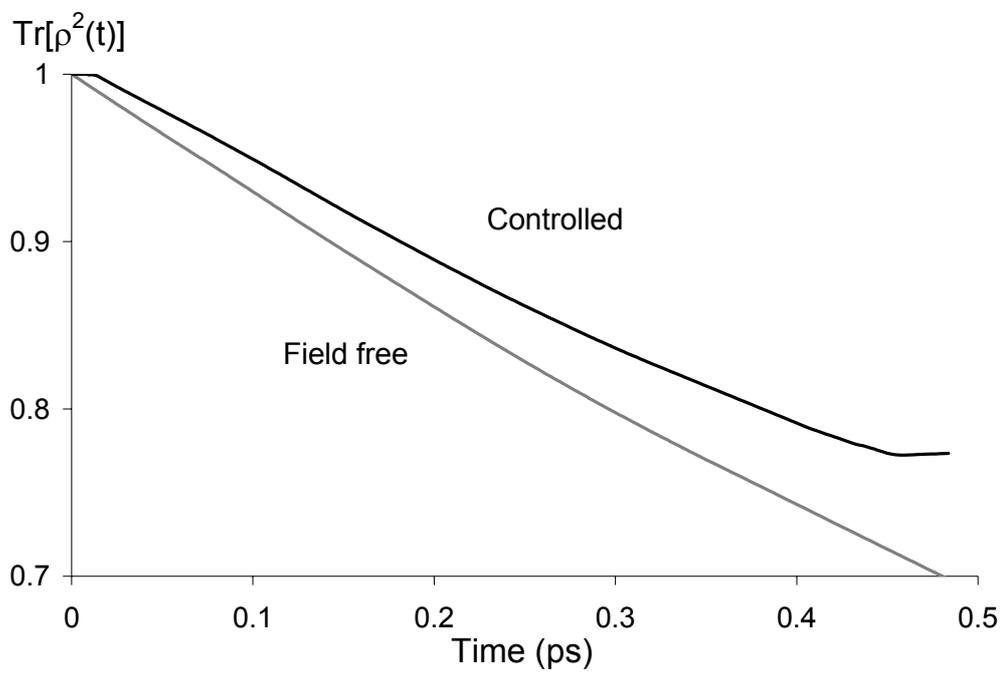

FIG. 7